\documentclass{article}

\usepackage{arxiv}

\usepackage[utf8]{inputenc} % allow utf-8 input
\usepackage[T1]{fontenc}    % use 8-bit T1 fonts
\usepackage{hyperref}       % hyperlinks
\usepackage{url}            % simple URL typesetting
\usepackage{booktabs}       % professional-quality tables
\usepackage{amsfonts}       % blackboard math symbols
\usepackage{nicefrac}       % compact symbols for 1/2, etc.
\usepackage{microtype}      % microtypography
\usepackage{lipsum}		% Can be removed after putting your text content
\usepackage{graphicx}
\usepackage[square , sort, numbers]{natbib}
\usepackage{doi}
\usepackage[nolist]{acronym}
\usepackage{threeparttable} %add by song
\usepackage{tabularx} %add by song
\usepackage{tabularx}
\usepackage{multirow}
\usepackage{threeparttable}

\title{An Experimental Study of Machine Learning-Based Intrusion Detection for OPC~UA over Industrial Private~5G Networks}

\date{March 24, 2026}	% Here you can change the date presented in the paper title
\author{ \href{https://orcid.org/0000-0003-1716-5867}{\includegraphics[scale=0.06]{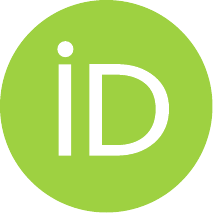}\hspace{1mm}Song Son Ha}\\
	Electrical Measurement Engineering\\
	Helmut-Schmidt-University\\
	Hamburg, Germany\\
	\texttt{song.ha@hsu-hh.de} \\
	\And
	\href{https://orcid.org/0009-0006-1601-2761}{\includegraphics[scale=0.06]{orcid.pdf}\hspace{1mm}Kunal Singh}\\
	Electrical Measurement Engineering\\
	Helmut-Schmidt-University\\
	Hamburg, Germany\\
	\texttt{kunal.singh@hsu-hh.de} \\
	\And
	\href{https://orcid.org/0009-0003-4437-010X}{\includegraphics[scale=0.06]{orcid.pdf}\hspace{1mm}Florian Foerster}\\
	Institute for Innovative Safety and Security\\
	Technical University of Applied Sciences Augsburg\\
	Augsburg, Germany\\
	\texttt{florian.foerster@tha.de} \\
	\And
	\href{https://orcid.org/0000-0002-5390-3946}{\includegraphics[scale=0.06]{orcid.pdf}\hspace{1mm}Henry Beuster}\\
	Electrical Measurement Engineering\\
	Helmut-Schmidt-University\\
	Hamburg, Germany\\
	\texttt{henry.beuster@hsu-hh.de} \\
	\And
	{\hspace{1mm}Tim Kittel}\\
	ipoque GmbH\\
	A Rohde \& Schwarz company\\
	Leipzig, Germany\\
	\texttt{tim.kittel@rohde-schwarz.com} \\
	\And
	\href{https://orcid.org/0000-0003-2310-5895}{\includegraphics[scale=0.06]{orcid.pdf}\hspace{1mm}Dominik Merli}\\
	Institute for Innovative Safety and Security\\
	Technical University of Applied Sciences Augsburg\\
	Augsburg, Germany\\
	\texttt{dominik.merli@tha.de} \\
	\And
	{\hspace{1mm}Gerd Scholl}\\
	Electrical Measurement Engineering\\
	Helmut-Schmidt-University\\
	Hamburg, Germany\\
	\texttt{gerd.scholl@hsu-hh.de} \\
}

\renewcommand{\shorttitle}{An Experimental Study of ML-Based IDS for OPC~UA over Industrial Private 5G~Networks}

\hypersetup{
	pdftitle={An Experimental Study of Machine Learning-Based Intrusion Detection for OPC~UA over Industrial Private~5G Networks},
	pdfsubject={OPC~UA, Intrusion Detection System, Private~5G, Machine Learning},
	pdfauthor={Song Son Ha},
	pdfkeywords={OPC~UA, Intrusion Detection System, Private~5G, Machine Learning},
}

\begin{document}
	\maketitle
	
	%\begin{acronym}
	%    \acro{iolw}[IOLW]{IO-Link Wireless}
	%    \acro{plc}[PLC]{programmable logic controller}
	%    \acro{sfrt}[SFRT]{safety function response time}
	%    \acro{ism}[ISM]{industrial, scientific and medical}
	%    \acro{mmtc}[mMTC]{massive machine type communications}
	%    \acro{iol}[IOL]{IO-Link}
	%    \acro{lte}[LTE]{long-term evolution}
	%    \acro{ue}[UE]{user equipment}
	%    \acro{nsa}[NSA]{non standalone}
	%    \acro{sa}[SA]{standalone}
	%    \acro{scs}[SCS]{subcarrier spacing}
	%    \acro{ofdm}[OFDM]{orthogonal frequency-division multiplexing}
	%    \acro{urllc}[URLLC]{ultra reliable low latency communications}
	%    \acro{embb}[eMBB]{enhanced mobile broadband}
	%    \acro{revpi}[RevPi]{Revolution Pi}
	%    \acro{vpn}[VPN]{virtual private network}
	%    \acro{dhcp}[DHCP]{Dynamic Host Configuration Protocol}
	%    \acro{ip}[IP]{Internet Protocol}
	%    \acro{rssi}[RSSI]{Received Signal Strength Indicator}
	%    \acro{iols}[IOLS]{IO-Link Safety}
	%    \acro{cpfen}[CPFEN]{Cyber Physical Finite Element Sensor Network}
	%    \acro{wmaster}[W-Master]{Wireless-Master}
	%\end{acronym}
	
	\begin{abstract}
		\footnote{This is the author's version of a paper that has been accepted for presentation at the 9th IEEE International Conference on Industrial Cyber-Physical Systems (ICPS 2026), to be held in Perth, Australia, on May 11–14, 2026.}
		Industrial deployments increasingly rely on Open Platform Communications Unified Architecture (OPC~UA) as a secure and platform-independent communication protocol, while private Fifth Generation (5G) networks provide low-latency and high-reliability connectivity for modern automation systems. However, their combination introduces new attack surfaces and traffic characteristics that remain insufficiently understood, particularly with respect to machine learning-based intrusion detection systems (ML-based IDS). This paper presents an experimental study on detecting cyberattacks against OPC~UA applications operating over an operational private~5G network. Multiple attack scenarios are executed, and OPC~UA traffic is captured and enriched with statistical flow-, packet-, and protocol-aware features. Several supervised ML models are trained and evaluated to distinguish benign and malicious traffic. The results demonstrate that the proposed ML-based IDS achieves high detection performance for a representative set of OPC~UA-specific attack scenarios over an operational private~5G network.
	\end{abstract}
	
	\acresetall
	
	%\keywords{OPC~UA, Intrusion Detection System, Private~5G, Machine Learning}
	
	\section{Introduction}
	
	The digital transformation of industrial systems is driving the adoption of communication technologies that combine interoperability, mobility, and stringent latency requirements. OPC~UA has become a dominant industrial communication standard due to its platform-independent design, object-oriented information modeling, and built-in security mechanisms \cite{Integration5GOPCUA}. In parallel, private~5G networks are increasingly deployed in industrial environments, offering low latency, high reliability, and scalable connectivity tailored to Industry~4.0 applications \cite{aijaz2020private,dutta20205g}. The integration of OPC~UA with private~5G is therefore expected to form a key communication backbone for future industrial automation systems.
	At the same time, this combination introduces new security challenges. While OPC~UA supports multiple security profiles, real-world industrial deployments often rely on mixed or unencrypted configurations for performance and compatibility reasons \cite{dahlmanns2020easing,Post2009}. Moreover, when OPC~UA traffic is transmitted over private~5G networks, wireless-specific effects such as dynamic scheduling and variable latency alter traffic timing characteristics compared to wired industrial networks, which may affect IDS designs typically developed under stable Ethernet conditions.
	Existing research on OPC~UA security predominantly focuses on wired deployments and a limited set of protocol-layer attacks \cite{BSI2017,vom2022opc,Neu2019,polgeassessing}. Conversely, most ML-based IDS research focuses on generic mobile traffic or public datasets that do not capture the semantics and operational characteristics of industrial protocols. Therefore, the behavior of OPC~UA traffic over private~5G networks and the effectiveness of ML-based IDS in such environments remain insufficiently explored.
	
	This paper addresses this research gap by deploying a realistic industrial testbed and executing multiple OPC~UA-specific attack scenarios targeting transport, session, and service layers. Traffic captured at the 5G user-plane interface is enriched with statistical flow-, packet-, and protocol-aware features, and several supervised ML models are trained and evaluated to distinguish benign and malicious traffic.
	
	The remainder of this paper is organized as follows. Related work is discussed in Section~\ref{chap:RelatedWork}, followed by background in Section~\ref{chap:Background}. The system architecture and threat model are presented in Section~\ref{chap:Architecture}, while the testbed, attack scenarios, feature extraction, and applied ML methods are described in Section~\ref{chap:Communication}. The evaluation results are presented in Section~\ref{chap:Evaluation}, followed by the conclusion and future work in Section~\ref{chap:Conclusions}.

	\section{Related Work} \label{chap:RelatedWork}

	Related work on securing OPC~UA and 5G-enabled industrial communication can be broadly categorized into vulnerability analyses, ML-based IDS using deep packet inspection (DPI), as well as security testbeds and datasets for 5G-enabled industrial control systems (ICS).

	Several studies analyze vulnerabilities and Denial-of-Service (DoS) threats in OPC~UA deployments, primarily focusing on conventional Ethernet-based environments. Neu et al.~\cite{Neu2019} investigate attacks by untrusted OPC~UA clients, while other work assesses the impact of protocol-layer attacks in Industry~4.0 scenarios using wired setups \cite{polgeassessing}. In addition, the German Federal Office for Information Security (BSI) provides extensive threat and vulnerability analyses of OPC~UA, concluding that while the security model mitigates certain risks, it does not prevent DoS attacks \cite{BSI2017,vom2022opc}. As a result, detection-based security mechanisms have been increasingly investigated.

	ML-based IDS using DPI have been widely studied in ICS and general network security. Prior work demonstrates that combining DPI with ML enables the detection of complex attacks and evasion strategies, motivating protocol-aware feature extraction \cite{Bindra2024,ResearchDPIML}. However, these approaches are typically evaluated on generic Information Technology (IT) or mobile traffic and do not model industrial protocols such as OPC~UA, nor do they consider the timing and scheduling dynamics introduced by private~5G networks.
	
	Research on 5G-enabled industrial environments increasingly explores ML-based security mechanisms and experimental testbeds. Lee et al.~\cite{Jonghoon} and Sharma et al.~\cite{sharmarole} discuss the integration of ML techniques into 5G industrial Internet of Things architectures, but without focusing on OPC~UA traffic or detailed protocol semantics. Several 5G security testbeds have been proposed to study attacks and traffic behavior \cite{VET5G,TestbedBaccar,almazyad}, yet they generally lack integrated ML-based IDS pipelines and do not specifically target OPC~UA-based industrial communication.

	Beyond protocol- and network-specific studies, multiple works emphasize the importance of realistic datasets and environments for evaluating IDS in ICS. Catillo et al.~\cite{Catillo2020} show that public DoS datasets may not generalize across different industrial systems due to device- and configuration-dependent traffic characteristics. Other studies highlight that IDS performance is strongly influenced by the realization of the evaluation environment and that IT-centric detection approaches often struggle with ICS-specific attacks \cite{Conti2021,Storm2024,Silaa2022}.

	In summary, there is still a lack of empirical studies that jointly consider OPC~UA semantics, real private~5G deployments, and ML-based intrusion detection. This paper addresses this gap by presenting a real-world setup over a commercial-grade private~5G network and systematically executing OPC~UA-specific attacks.

	\section{Background} \label{chap:Background}
	
	\subsection{ICS Security}
	
	ICS combine Operational Technology (OT), which monitors and controls physical processes, with IT components for data processing and enterprise connectivity. Typical OT environments comprise field devices and sensors, Remote Terminal Units for signal acquisition, Programmable Logic Controllers (PLCs) for automation and real-time control, industrial networks, and Human-Machine Interfaces (HMIs) for operator supervision and intervention \cite{Kumar2025}. A key security distinction is the different priority order: IT systems are usually designed around privacy and confidentiality, authentication, integrity, and non-repudiation, whereas OT systems prioritize availability first, followed by integrity and confidentiality, since loss of availability can directly interrupt production \cite{Garimella2018}. Historically, OT networks were isolated and relied on proprietary stacks, but Industry 4.0 has driven convergence with standard IT technologies such as Ethernet, IP, and wireless links. While this increases interoperability and remote access, it reduces isolation and enlarges the attack surface, making ICS more exposed to network-based threats \cite{Stouffer2023}. Conducting security research on productive industrial plants is risky because even minor disturbances can cause costly incidents, so controlled security testbeds are used to reproduce industrial conditions and evaluate attacks and defenses without endangering real infrastructure \cite{Sauer2019,Christiansson}.

	\subsection{OPC~UA}
	
	OPC~UA is a dominant industrial communication standard designed for heterogeneous ICS environments. Its platform-independent architecture and object-oriented information modeling enable interoperable representation of devices, variables, and services across vendors and automation layers \cite{Integration5GOPCUA}. OPC~UA defines structured interactions including endpoint discovery, secure channel establishment and session creation. It supports multiple underlying protocols, including TCP/IP, HTTPS, MQTT, and UDP-based variants, providing flexibility for various industrial requirements such as real-time performance and secure remote access. Security is embedded via selectable profiles providing authentication, integrity, and encryption, but real deployments may use mixed profiles or unencrypted modes for performance or compatibility, which weakens effective protection \cite{BSI2017,vom2022opc}. Because OPC~UA operates at the application layer with rich semantics and stateful sessions, it exposes attack surfaces beyond simple packet floods, motivating protocol-aware defenses and detection \cite{Neu2019,polgeassessing,bu2020}.

	\subsection{Private~5G Networks and Security Challenge}

	Private~5G networks, also known as campus or non-public networks, represent a paradigm shift in industrial connectivity. 
	5G was initially specified in 3GPP Release~15, with additional enhancements for industrial use cases introduced in Release~16~\cite{3gppTR21916Version}, and has since evolved through subsequent releases, currently reaching Release~19. Unlike previous cellular generations, 5G favors a standalone architecture that operates independently of 4G infrastructure and utilizes dedicated spectrum to support high-bandwidth and deterministic applications. The 5G architecture relies on softwarization principles to efficiently and flexibly manage complex networks and enable the dynamic provisioning of resources, such as ultra-reliable low-latency communication, to meet specific industrial requirements.
	At the same time, the transition from wired fieldbus systems to software-defined wireless IP-based networks fundamentally alters the security challenge. The inherent openness of the Radio Access Network introduces physical layer vulnerabilities, making critical control traffic susceptible to jamming, signal spoofing, and eavesdropping \cite{lichtman5GNRJamming2018}. Furthermore, the shift toward logical rather than physical isolation 
	achieved through network slicing expands the attack surface. If logical separation measures in the virtualized core are compromised, it introduces risks of lateral movement and inter-slice side-channel attacks, challenging the traditional perimeter-based defense models used in OT environments.

	\subsection{IDS and ML}
	
	IDS are commonly classified into signature-based and anomaly-based approaches. Signature-based IDS detect intrusions by matching observed activity against known attack patterns, achieving low false-positive rates for previously identified threats but remaining ineffective against novel or evolving attacks without frequent signature updates \cite{Kumar2025,Stouffer2023}. Anomaly-based IDS model normal system or network behavior and flag deviations as potential intrusions. This approach is particularly suitable for ICS, where communication patterns are typically repetitive and deterministic during normal operation. However, anomaly-based detection is sensitive to concept drift, which may increase false positives if not carefully designed \cite{Kumar2025,Conti2021}. 
	ML techniques have been widely adopted to enhance anomaly-based IDS by enabling automated pattern learning, analysis of high-dimensional traffic features, and the detection of complex attack behaviors that are difficult to capture using rule-based approaches. In industrial environments, ML-based IDS can leverage temporal, statistical, and protocol-aware features to model normal system behavior and identify subtle deviations. Recent studies highlight the applicability of ML-based IDS in 5G-enabled industrial networks, where dynamic traffic characteristics and increased system complexity challenge traditional detection mechanisms \cite{Jonghoon,sharmarole}.
	
	For industrial application-layer protocols such as OPC~UA, traditional packet- or flow-level IDS are often insufficient, as attacks may exploit stateful interactions, service semantics, and session or secure channel management rather than traffic volume alone. When OPC~UA communication is deployed over private~5G networks, challenges arise due to altered timing, scheduling, and multiplexing characteristics compared to wired OT environments, which can influence traffic features even under benign conditions. These aspects motivate protocol-aware feature extraction and experimental evaluation of ML-based IDS to assess detection performance and feature robustness for OPC~UA over industrial private~5G networks.

	\section{Concept and Architecture} \label{chap:Architecture}
	
	\subsection{Threat Model}

	We assume basic as well as advanced adversaries that can act both as an external or an internal attacker with direct access to the ICS network. In particular, we consider that in wireless or shared infrastructures it is difficult to distinguish between external and internal origins of traffic due to the wireless nature, and that an attacker may also compromise ICS components such as PLCs or other field devices to launch attacks in the network.

	Under the experimental threat model, unencrypted OPC~UA traffic is considered to ensure transparency, reproducibility, and controlled experimentation. Accordingly, \texttt{SignAndEncrypt} security modes are disabled, increasing the exposed attack surface compared to end-to-end protected deployments. An ML-based IDS with DPI capabilities is assumed as a compensating monitoring and detection mechanism under these experimental conditions.

	While the testbed supports multiple communication protocols, the presented work concentrates on OPC~UA, as related work on 5G and IDS integration has already covered other protocols more extensively but does not consider OPC~UA-specific behavior and service semantics \cite{Jonghoon,VET5G,almazyad,TestbedBaccar}. The implemented attacks are representative OPC~UA-specific denial-of-service and service abuse scenarios designed to stress different protocol dimensions, including transport behavior, session and channel management, service invocation dynamics, and semantic parsing complexity. However, they do not aim to exhaustively cover all possible attack classes.

	\subsection{Architecture Design}
	
	The proposed architecture builds upon a conceptual design introduced in \cite{11205743} and extends it into a fully operational, OPC~UA-centric security evaluation pipeline deployed over a private~5G network. It supports larger deployments, multiple attack types, multi-layer feature extraction, and end-to-end ML evaluation. 
	Within this setup, OPC~UA clients and servers operate in a dedicated application cell, where operational process values are exchanged, while a separate attack cell provides a controlled environment for generating malicious traffic representative of realistic cyber threats. All benign and adversarial traffic is mirrored and forwarded to a DPI component, which parses OPC~UA protocol messages and extracts protocol-aware as well as network-level features.
	The extracted features are aggregated and processed into a labeled dataset, subsequently split into training, validation, and test sets for training and evaluating supervised ML models that form the core of the ML-based IDS for classifying benign and malicious OPC~UA traffic.

	\section{Testbed Implementation} \label{chap:Communication}

	\subsection{Hardware Components}
	
	The experimental testbed is deployed on an operational private~5G Standalone network using dedicated industrial and server-grade hardware. The attack cell is hosted on a dedicated Dell Precision 3630 workstation, generating malicious OPC~UA traffic. The application cell is deployed on industrial Revolution~Pi (RevPi) PLCs, which are connected to the private~5G network using 5G HAT modems equipped with SIMcom SIM8202G-M2 multi-band modules.
	
	The private~5G network is based on the Ericsson \mbox{EDAV-I} solution compliant with 3GPP Release~16, operating in the \mbox{3.7--3.8\,GHz} band and providing a mirrored user-plane interface with up to 10\,Gb/s throughput. Intrusion detection and traffic analysis are performed on a dedicated IDS cluster consisting of Dell R750xs and R660xs servers, leveraging the R\&S®PACE 2 library for real-time protocol analysis and feature extraction.

	\subsection{OPC~UA-based Application}

	\begin{figure}[t]
		\centering
		\includegraphics[width=0.6\linewidth]{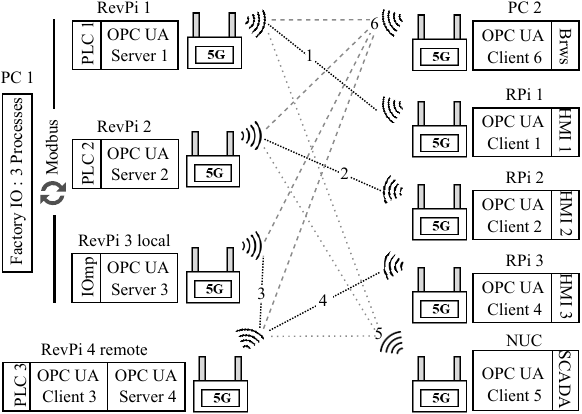}
		\caption{Architecture of the application cell in 5G industrial testbed.}
		\label{fig:architecture_combined}
	\end{figure}

	This work implements an OPC~UA-based industrial application operating over a private~5G network and models a representative automated warehouse using Factory~I/O~\cite{factoryio}, a commercial industrial simulation platform for emulating sensor-actuator interactions in PLC-based automation scenarios. An overview of the OPC~UA-based application architecture is shown in Fig.~\ref{fig:architecture_combined}. The virtual warehouse comprises three concurrent processes associated with separate storage racks, each controlled by a PLC program capable of executing interchangeable placement strategies, including sequential, even-odd, and random slot selection. Two PLC programs run locally on RevPi~1 and RevPi~2, while the third runs remotely on RevPi~4 over the 5G network. Material handling is realized using simulated conveyors, sensors, and actuators, with process I/O exchanged via Modbus between Factory~I/O and RevPi~1, RevPi~2, and RevPi~3.
	Based on this application setup, several OPC~UA communication paths are defined. Paths~1, 2, and~4 correspond to HMI-based supervision, where HMI clients connect to OPC~UA Servers~1, 2, and~4 to monitor process states and issue operator commands. Path~3 represents remote PLC control over 5G, where PLC logic running on RevPi~4 uses OPC~UA Client~3 to interact with OPC~UA Server~3 on RevPi~3 and maps selected variables to OPC~UA Server~4 for HMI access. Path~5 denotes SCADA-level monitoring via a NUC-based workstation, while Path~6 corresponds to a non-periodic OPC~UA browsing client generating traffic representative of engineering and diagnostic tools.

	\begin{table}[t]
		\centering
		\caption{Selected OPC~UA Traffic Scenarios and Applied Parameters$^{\mathrm{*}}$}
		\renewcommand{\arraystretch}{1.05}
		\begin{threeparttable}
			\begin{tabularx}{\textwidth}{@{}l l X@{}}
				\toprule
				\textbf{Grp.} & \textbf{Traffic Scenario} & \textbf{Applied Parameters} \\
				\midrule
				
				\multirow{2}{*}{G0} &
				Benign (Normal Operation) &
				Benign-only OPC~UA traffic representing steady and variable operational processes. \\
				
				\midrule
				\multirow{7}{*}{G1} &
				
				HEL Message Flooding $^{\dagger}$ (HEL-F) &
				High-rate \texttt{HEL} request messages at target rates of 500, 1000, and 2000 HEL/s. \\
				
				& Open Multiple Secure Channels (OMSC) &
				Multiple \texttt{SecureChannel} openings within a single session at rates of 50, 100, and 150 channels/s. \\
				
				& Chunked Message Flooding (CHUNK-F) &
				Chunk-flood attempts using oversized OPC~UA messages (about 200-300\,kB) split into chunks of 200, 1000, and 4000 bytes; the final chunk is intentionally omitted. \\
				
				\midrule
				\multirow{7}{*}{G2} &
				
				Publish Request Flooding$^{\dagger}$ (PUB-F) &
				Repeated \texttt{Publish} requests at rates of 50, 100, and 150 requests/s. \\
				
				& Unlimited Condition Refresh (COND-REF) &
				Repeated \texttt{ConditionRefresh} calls on active subscriptions at rates of 30, 50, and 100 requests/s. \\
				
				& Browse Address Space$^{\dagger}$ (BROWSE) &
				Iterative \texttt{Browse} operations starting from \texttt{RootFolder} with randomized shallow and deep traversals at rates of 20, 40, and 80 requests/s. \\
				
				\midrule
				\multirow{9}{*}{G3} &
				
				Read Request List Expansion$^{\dagger}$ (READ-EXP) &
				Low-rate \texttt{Read} requests at 5 requests/s with increasing \texttt{NodesToRead} list sizes of 16, 32, and 64 to isolate payload complexity effects. \\
				
				& Complex Nested Message (NESTED) &
				Low-rate \texttt{Write} requests at 5 requests/s with increasing \texttt{Variant} nesting depths of 16, 32, and 64 to isolate parsing and deserialization overhead. \\
				
				& Translate Browse Path (TBP) &
				\texttt{TranslateBrowsePathsToNodeIds} requests at 20 requests/s with \texttt{RelativePath} depths of 32, 64, and 128 to isolate semantic complexity. \\
				
				\bottomrule
			\end{tabularx}
			
			\begin{tablenotes}[flushleft,para]
				\footnotesize
				\item[$^{\mathrm{*}}$]
				G0: benign baseline traffic;
				G1: transport- and session-level flooding;
				G2: service invocation flooding;
				G3: complexity-driven service abuse.
				All attack scenarios are executed using a periodic burst-idle pattern with a 3\,s burst followed by a 7\,s idle phase.
				
				\item[$^{\dagger}$]
				Implemented as extensions of Claroty’s opcua-exploitation framework; other scenarios rely on parameterized configurations.
			\end{tablenotes}
		\end{threeparttable}
		\label{tab:attacks}
	\end{table}

	\subsection{Attack Selection and Deployment}
	
	Based on the defined threat model, the evaluation employs a focused set of OPC~UA-specific attack scenarios built upon Claroty’s opcua-exploitation framework~\cite{claroty2025}, using parameterized configurations and targeted extensions. The attacks, summarized in Table~\ref{tab:attacks}, are grouped into transport- and session-level flooding, service invocation flooding, and complexity-driven service abuse. Despite differing execution characteristics, all attack scenarios ultimately aim at exhausting OPC~UA server resources through either high-rate flooding or low-rate semantic and parsing complexity.

	While the number of attack types is limited, each scenario is executed under multiple parameterizations and durations, resulting in significantly different temporal, volumetric, and semantic traffic characteristics within the same attack class. For each attack scenario, three parameter configurations were evaluated, each executed in three runs of approximately 3, 6, and 9~minutes, resulting in about 54~minutes of captured traffic per attack, covering short and longer attack durations. The collected packet capture (PCAP) files contain either pure benign traffic or a mixture of benign and attack-induced traffic, including both warm-up and recovery phases. 
	Ten benign-only PCAP files were recorded, each lasting approximately 30~minutes and obtained from repeated executions of the same application setup with varying process behaviors.

	\begin{table}[t]
		\centering
		\caption{Flow-derived Base Features}
		\renewcommand{\arraystretch}{1.05}
		\begin{threeparttable}
			\begin{tabular}{@{}p{0.28\columnwidth} p{0.4\columnwidth}@{}}
				\toprule
				\textbf{Feature / Group} & \textbf{Description} \\
				\midrule
				
				Identifier &
				flow\_id, window\_id, server/client IPs and ports \\
				
				\midrule
				\multicolumn{2}{l}{\textit{Transport-level temporal and size features}} \\
				
				t\_pkt\_count & Total packet counts \\
				t\_pkt\_size\_sum & Aggregated packet size in total \\
				t\_body\_\{count,length\_sum\} &
				OPC~UA message body count and length \\
				t\_\{fwd,bwd\}\_pkt\_count & Forward and backward packet counts \\
				t\_\{fwd,bwd\}\_pkt\_size\_sum & Forward and backward packet size in total \\
				
				t\_iat\_\{sum,count\} &
				Packet inter-arrival time sum and count \\
				
				t\_active\_flow\_drt & Flow active duration \\
				
				\midrule
				\multicolumn{2}{l}{\textit{OPC~UA service-level aggregate features}} \\
				t\_sec\_ch\_count & Counts \texttt{SecureChannelId} transitions \\
				t\_service\_\{req,res\}\_count & OPC~UA service requests and responses count\\
				t\_server\_res\_time\_sum & OPC~UA server response latency \\
				t\_server\_res\_time\_count & OPC~UA server response latency count\\
				
				\midrule
				\multicolumn{2}{l}{\textit{OPC~UA service-specific semantic features}} \\
				
				t\_\{read,write\}\_\{req,res\} &
				Read and Write data access \\
				
				t\_publish\_\{req,res\} &
				Publish notification service \\
				
				t\_open\_sec\_ch\_\{req,res\} &
				\texttt{SecureChannel} establishment \\
				
				t\_close\_sec\_ch\_\{req,res\} &
				\texttt{SecureChannel} termination \\
				
				t\_create\_session\_\{req,res\} &
				Session creation \\
				
				t\_activate\_session\_\{req,res\} &
				Session activation \\
				
				t\_close\_session\_\{req,res\} &
				Session closure \\
				
				t\_create\_mnt\_itm\_\{req,res\} &
				\texttt{MonitoredItems} creation \\
				
				t\_\{create,del\}\_sub\_\{req,res\} &
				Subscription creation and deletion \\
				
				t\_endpoints\_\{req,res\} &
				Endpoint discovery \\
				
				\bottomrule
			\end{tabular}
			
		\end{threeparttable}
		\label{tab:features1}
	\end{table}

	\subsection{Feature Extraction and Processing}
	\label{subsec:feature_extraction}
	
	OPC~UA traffic features are extracted over fixed, non-overlapping time windows (TWs) of 5\,s using a two-stage aggregation approach. Statistics are first computed at the flow level and then aggregated across all flows within each TW, enabling the capture of both traffic intensity characteristics and protocol-level behavior.

	\begin{table}[t]
		\centering
		\caption{Derived Cross-flow Window-level Features}
		\renewcommand{\arraystretch}{1.05}
		\begin{threeparttable}
			\begin{tabular}{@{}p{0.28\columnwidth} p{0.64\columnwidth}@{}}
				\toprule
				\textbf{Feature / Group} & \textbf{Description} \\
				\midrule
				
				\multicolumn{2}{l}{\textit{Ratio- and mean-based aggregations}} \\
				
				gl\_mean\_pkt\_size &
				Mean packet size \\
				
				gl\_mean\_body\_len &
				Mean OPC~UA body length \\
				
				gl\_body\_pkt\_ratio &
				Fraction of packets carrying OPC~UA payload over all packets \\
				
				gl\_body\_byte\_ratio &
				Fraction of OPC~UA payload bytes over total packet bytes \\
				
				gl\_fwd\_pkt\_ratio &
				Fraction of forward packets \\
				
				gl\_fwd\_byte\_ratio &
				Fraction of bytes attributed to forward packets \\

				gl\_mean\_iat &
				Mean packet inter-arrival time \\
				
				gl\_\{mean,max\}\_flow\_drt &
				Mean and maximum flow duration \\
				
				gl\_req\_res\_ratio &
				Ratio of service requests to responses \\
				
				gl\_mean\_server\_res\_time &
				Mean server-side response latency \\
				
				gl\_req\_per\_flow &
				Mean number of service requests per flow \\
				
				gl\_\{read,write,publish\}\_ratio &
				Fractions of Read, Write, and Publish requests among all service requests \\
				
				gl\_monitored\_item\_rate &
				Mean number of \texttt{CreateMonitoredItems} requests per flow \\
				
				\midrule
				\multicolumn{2}{l}{\textit{Count- and event-based aggregations}} \\
				
				gl\_flow\_count &
				Number of active flows \\
				
				gl\_sec\_ch\_churn &
				\texttt{SecureChannel} open and close activity \\
				
				gl\_session\_churn &
				Session lifecycle activity \\
				
				gl\_sub\_churn &
				Subscription create and delete activity \\
				
				\midrule
				\multicolumn{2}{l}{\textit{Flow dominance and service diversity features}} \\
				
				gl\_max\_flow\_pkt\_ratio &
				Maximum fraction of packets contributed by a single flow \\
				
				gl\_max\_flow\_req\_ratio &
				Maximum fraction of OPC~UA service requests contributed by a single flow \\
				
				gl\_max\_flow\_publish\_ratio &
				Maximum fraction of Publish requests contributed by a single flow \\
				
				gl\_std\_flow\_req\_count &
				Standard deviation of service request counts across flows \\
				
				gl\_std\_flow\_publish\_count &
				Standard deviation of Publish request counts across flows \\
				
				gl\_service\_entropy &
				Shannon entropy of OPC~UA service request distribution \\
				
				\bottomrule
			\end{tabular}

		\end{threeparttable}
		\label{tab:features2}
	\end{table}

	\subsubsection{Flow-derived base features}
	As a first step, transport-level and OPC~UA service-level statistics are computed for each bidirectional flow observed within a TW. These base features, denoted by the prefix \texttt{t\_} and summarized in Table~\ref{tab:features1}, include packet counts and sizes, inter-arrival time statistics, flow active duration, secure channel and session lifecycle events, as well as service-specific request and response counters and subscription-related services. 
	
	Although initially extracted at the flow level, the \texttt{t\_} features are subsequently aggregated across all flows within the same TW using simple statistical operators such as sum, mean, maximum, and count. The resulting window-level aggregates form the primary statistical inputs to the IDS and provide a compact representation of overall traffic activity, protocol usage, and lifecycle activity.

	\subsubsection{Derived cross-flow features}
	Based on the window-level aggregation of the flow-derived \texttt{t\_} statistics, a second set of cross-flow features is constructed to characterize the overall traffic composition, interaction patterns, and structural dynamics within each TW. These features, denoted by the prefix \texttt{gl\_} and summarized in Table~\ref{tab:features2}, are designed to complement the aggregated base statistics by capturing relative, collective, and higher-order properties that are not observable at the individual flow level.
	
	Ratio- and mean-based features describe the normalized distribution of traffic volume, timing, and service activity across flows, enabling consistent characterization under varying load conditions, independent of absolute traffic rates. Count- and event-based features capture the intensity and recurrence of protocol lifecycle operations, such as SecureChannel, Session, and Subscription management, reflecting state-related dynamics and interaction frequency rather than raw packet volume alone. In addition, flow dominance and service diversity features quantify how traffic and service usage are distributed across concurrent flows, providing sensitivity to imbalances, concentration effects, and shifts in protocol usage patterns.
	
	Accordingly, the derived cross-flow features emphasize relative and structural characteristics of OPC~UA communication that are less dependent on specific attack signatures, supporting robust detection under varying benign workloads and private~5G-induced timing variability.
	
	Together, the aggregated \texttt{t\_} features and the derived \texttt{gl\_} features constitute the final IDS feature vector used for model training and evaluation.

	\subsection{Applied ML Algorithms}
	
	To assess the effectiveness of ML for intrusion detection in OPC~UA traffic over a private~5G network, a diverse set of supervised learning models was evaluated, covering different learning paradigms commonly used in IDS research. Logistic Regression (LogReg) was included as a linear baseline to evaluate how well benign and malicious OPC~UA traffic can be separated using the extracted features. Support Vector Machines (SVM) with radial basis function kernels were considered to capture non-linear relationships between features and attack behaviors. Random Forest (RF) was selected as a representative bagging-based ensemble method due to its robustness across heterogeneous traffic patterns. Gradient Boosting (GB) and Extreme Gradient Boosting (XGBoost) were applied as boosting-based approaches that iteratively refine classification performance by learning from previous prediction errors. In addition, a soft Voting Ensemble (Voting) combining Random Forest, Logistic Regression, and k-Nearest Neighbors was evaluated to examine whether aggregating complementary classifiers improves robustness across diverse attack scenarios.
	
	The dataset was partitioned into three disjoint subsets for training, validation, and testing using a PCAP-level separation strategy, meaning that all traffic windows originating from the same packet capture file were assigned exclusively to a single split. This prevents traffic leakage between splits and ensures that temporal correlations within individual traces do not artificially inflate detection performance. The resulting splits comprise 61 PCAPs for training, 10 PCAPs for validation, and 20 PCAPs for testing. Depending on the capture, a PCAP may contain benign-only traffic or a combination of benign and attack-induced OPC~UA traffic with different attack types and parameter severities.
	
	The training split was used to fit model parameters, while the validation split was exclusively employed for hyperparameter selection in models supporting grid-search optimization. Final performance evaluation was conducted only on the held-out test split. Detection performance is reported using class-wise F1-scores, which capture the balance between false alarms and missed detections and are particularly suitable for intrusion detection under class imbalance and heterogeneous attack characteristics.

	\section{Performance Evaluation Results}
	\label{chap:Evaluation}
	This section presents the performance evaluation of the proposed ML-based IDS using the extracted OPC~UA traffic features. The discussion focuses on detection performance across different traffic scenarios, model robustness across learning paradigms, and the implications of deploying the IDS under realistic private~5G operating conditions, with the class-wise F1-scores of the evaluated ML models on the test set reported in Table~\ref{tab:Performance}.

	\subsection{Detection Performance by Traffic Scenario}
	
	To interpret the classification results, detection performance is analyzed according to the traffic scenarios defined in Table~\ref{tab:attacks}, which represent benign baseline operation as well as adversarial strategies at different layers of the OPC~UA protocol.

	\begin{table}[t]
		\centering
		\caption{F1-Score Comparison Across All Evaluated Models (Test Set)}
		\footnotesize
		\setlength{\tabcolsep}{6.5pt}
		\renewcommand{\arraystretch}{1.15}
		\begin{tabular}{l|cccccc}
			\toprule
			Attack & LogReg & SVM & RF & GB & XGBoost & Voting \\
			\midrule
			BENIGN   
			& 0.976 & 0.976 & 0.976 & \textbf{0.986} & \textbf{0.986} & 0.981 \\
			\midrule
			
			HEL-F    
			& \textbf{1.000} & 0.995 & \textbf{1.000} & \textbf{1.000} & \textbf{1.000} & \textbf{1.000} \\
			
			OMSC     
			& \textbf{0.976} & 0.970 & 0.970 & 0.970 & \textbf{0.976} & \textbf{0.976} \\
			
			CHUNK-F  
			& \textbf{0.960} & 0.946 & 0.846 & 0.946 & 0.917 & 0.935 \\
			
			\midrule
			PUB-F    
			& 0.974 & 0.965 & 0.974 & 0.987 & \textbf{0.996} & 0.991 \\
			
			COND-REF 
			& 0.962 & 0.976 & \textbf{0.986} & 0.954 & 0.978 & 0.973 \\
			
			BROWSE   
			& 0.938 & 0.958 & 0.923 & 0.956 & \textbf{0.971} & 0.963 \\
			
			\midrule
			READ-EXP 
			& 0.868 & 0.918 & 0.920 & \textbf{0.961} & 0.954 & 0.957 \\
			
			NESTED   
			& 0.896 & 0.907 & 0.753 & \textbf{0.915} & \textbf{0.915} & 0.912 \\
			
			TBP      
			& 0.934 & 0.929 & 0.879 & 0.924 & 0.933 & \textbf{0.935} \\
			
			\bottomrule
		\end{tabular}
		\label{tab:Performance}
	\end{table}	
	
	\subsubsection{Benign Baseline Traffic}
	
	Benign baseline traffic is detected with very high accuracy across all evaluated models, achieving F1-scores between approximately 0.97 and 0.99. This demonstrates that the normal operational behavior of the OPC~UA-based industrial application is classified reliably, even under varying process dynamics and the timing variability introduced by the private~5G network. The consistently high performance further indicates a low false-alarm tendency, which is critical for deployment in operational industrial environments.
	
	\subsubsection{Transport- and Session-level Flooding}
	
	Transport- and session-level flooding traffic scenarios are detected with strong performance across the evaluated models. These scenarios introduce pronounced volumetric and state-related deviations, such as excessive handshake initiation and repeated secure-channel establishment, which are effectively captured by the proposed feature set. 
	Nevertheless, moderate performance variations are observed for scenarios that rely on application-layer message structure rather than increased traffic volume. In particular, abnormal OPC~UA message chunking introduces only weak statistical deviations from benign behavior, making such scenarios more difficult to detect for some classifiers.

	\subsubsection{Service Invocation Flooding}
	
	Service invocation flooding traffic scenarios achieve consistently high detection performance across models. These scenarios generate abnormal service request dynamics and lifecycle activity, which are well reflected in the extracted protocol-aware service-level and cross-flow features.
	The stable detection performance observed across different learning paradigms confirms that service-level statistics provide robust indicators of abnormal behavior beyond simple traffic volume metrics, even when traffic patterns are executed using burst-idle strategies that partially mimic benign operational behavior.
	
	\subsubsection{Complexity-driven Service Abuse}
	
	Complexity-driven service abuse represents the most challenging traffic scenario category. These scenarios operate at low request rates, exploiting semantic and parsing complexity rather than volumetric anomalies, resulting in weaker statistical separation from benign traffic.
	While detection performance remains high for models with higher representational capacity, linear and bagging-based classifiers exhibit noticeable performance degradation for this traffic category. This behavior highlights the inherent difficulty of capturing short-lived semantic abuse within fixed aggregation windows and emphasizes the importance of expressive models for detecting complexity-driven traffic patterns in industrial protocols.
	
	\subsection{Model Comparison and Robustness}
	
	Comparing the evaluated classifiers, boosting-based models provide the most stable detection performance for complex traffic scenarios, particularly for complexity-driven service abuse. These scenarios operate at low request rates and primarily manifest through semantic and structural message characteristics, which are captured more consistently by models with higher representational capacity.
	
	In contrast, transport- and session-level flooding scenarios that produce strong volumetric increases or abnormal connection and channel handling behavior are detected reliably by most evaluated models. This indicates that such traffic patterns generate clear statistical deviations that can be identified even by simpler linear and ensemble-based classifiers.
	
	Service invocation flooding scenarios show little sensitivity to the choice of learning paradigm, with consistently high detection performance observed across all evaluated models. This suggests that protocol-aware service-level features are sufficient to characterize abnormal service usage patterns, making detection robustness less dependent on model complexity than for complexity-driven attacks.
	
	Overall, these findings align with prior work reporting that IDS approaches relying primarily on traffic volume or simple threshold-based indicators struggle to detect low-rate and semantics-driven behavior in industrial communication protocols. The results demonstrate that protocol-aware feature extraction, combined with learning models capable of capturing complex traffic patterns, is essential for robust intrusion detection in OPC~UA environments.

	\subsection{Discussion on Deployment in Private~5G Networks}
	
	The evaluation is conducted using OPC~UA traffic captured in a production-grade industrial private~5G network. Compared to wired industrial Ethernet, private~5G introduces additional timing variability due to radio scheduling and dynamic resource allocation.
	Despite this variability, high detection performance is maintained across all evaluated traffic scenarios. This indicates that the proposed IDS is not overly sensitive to transient wireless-induced timing fluctuations introduced by private~5G communication. Rather than relying on absolute timing information or fixed inter-packet thresholds, the evaluated models learn characteristic traffic patterns from protocol-aware statistics that combine flow-level behavior with aggregated cross-flow features capturing relative, structural, and service-level dynamics. As a result, the IDS remains robust to short-term packet timing variations and generalizes effectively under private~5G operating conditions.
	
	Overall, the results suggest that private~5G does not inherently hinder ML-based IDS for OPC~UA applications when appropriate feature design and aggregation strategies are applied, supporting the feasibility of deploying protocol-aware IDS solutions in industrial private~5G networks.

	\section{Conclusions and Future Work}
	\label{chap:Conclusions}
	
	This paper presented an experimental study on ML-based intrusion detection for OPC~UA applications operating over an industrial private~5G network. A testbed was implemented that is representative of industrial deployment scenarios integrating an OPC~UA-based industrial application, a commercial private~5G Standalone deployment, and a protocol-aware IDS pipeline based on DPI and multi-level traffic feature aggregation. A representative set of OPC~UA-specific attack scenarios targeting transport-, session-, and service-layer behavior was systematically executed and evaluated.		
	The evaluation results show that supervised ML models can reliably distinguish benign and malicious OPC~UA traffic across a diverse set of attack scenarios. High detection performance is achieved not only for transport- and session-level flooding attacks, but also for service invocation flooding and complexity-driven service abuse attacks that operate at low request rates. These results confirm that protocol-aware OPC~UA features capturing service usage, lifecycle activity, and semantic structure provide robust indicators of abnormal behavior when aggregated across flows within fixed TWs.
	Moreover, the results indicate that the timing variability introduced by private~5G communication does not inherently degrade IDS effectiveness. By relying on protocol-aware statistics derived from flow-level behavior and aggregated cross-flow dynamics rather than fixed timing thresholds, the proposed IDS generalizes well under wireless operating conditions, supporting its applicability in industrial private~5G environments.
	
	Future work will extend the evaluation toward encrypted OPC~UA communication, investigate feature transferability between wired Ethernet and private~5G deployments, and explore adaptive learning techniques to address long-term traffic evolution and concept drift.

	\section*{Acknowledgment}
	
	The authors would like to thank ipoque GmbH, J. Jockram and F. Mueller for their valuable support. We also gratefully acknowledge Deutsche Telekom and Ericsson for their continuous support and for accommodating requests that sometimes went beyond those of typical industrial customers. 
	
	\section*{Funding}
	This research is funded by dtec.bw – Digitalization and Technology Research Center of the Bundeswehr. dtec.bw is funded by the European Union – NextGenerationEU (project “Digital Sensor-2-Cloud Campus Platform” (DS2CCP), \href{https://dtecbw.de/home/forschung/hsu/projekt-ds2ccp}{https://dtecbw.de/home/forschung/hsu/projekt-ds2ccp}).
	
	\bibliographystyle{unsrtnat}
	\bibliography{references}
	
\end{document}